\documentstyle[tighten,prl,aps,multicol,epsfig]{revtex}

\begin{document}

\title{Entanglement Manipulation and Concentration}
\author{R.~T.~Thew\cite{RTT} and W. J. Munro\cite{WJM}}
\address{Special Research Centre for Quantum Computer Technology,\\
University of Queensland,  Brisbane, Australia}
\date{\today}

\maketitle
\begin{abstract}
We introduce a simple, experimentally realisable, entanglement
manipulation protocol for exploring mixed state entanglement. We show
that for both non-maximally entangled pure, and mixed 
polarisation-entangled two qubit states, an increase in the degree of
 entanglement and purity, which we define as concentration, is achievable. 

\end{abstract}


\label{INTRO}

The increasing interest in quantum information and computing as well as
other quantum mechanical dependent operations such as teleportation
\cite{Bennett2:96} and cryptography \cite{Deutsch:96} have as their
cornerstone a reliance on entanglement. There has been a great deal of 
discussion of measures and manipulation of  entanglement in recent years with
respect to  purification \cite{Bennett1:96}, concentration
\cite{Bennett3:96}, and distillable entanglement 
\cite{Rains1:99,Vedral:97} especially concerning states
subject to environmental noise. It is this noise that takes the initially
pure maximally entangled resource and leaves us with, at best, a non maximally
entangled state, or at worst a mixed state, both less pure and less entangled.
We introduce a simple, experimentally realisable \cite{Kwiat:00}, protocol to
manipulate and explore both pure and mixed-state entanglement. While
the scheme will have limitations, in part due to its simplicity, it
will allow experimental investigation of the large Hilbert space
associated with mixed states.  

The motivation for this scheme comes from focusing ideas and
proposals of several groups from the past few years into a
simple realisation of mixed state entanglement
manipulation. It was  proposed that quantum correlations on mixed
states could be enhanced by positive operator valued measurements 
\cite{Popescu1:95}. A more specific example by Gisin\cite{Gisin1:96}
considered  the manipulation of a $ 2\times 2 $ system using local 
filters. The scheme we propose here combines these ideas and uses 
an arrangement similar to the original Procrustean method \cite{Bennett3:96} 
which dealt solely with pure states. The primary motivation here is in 
proposing a scheme that can be easily realised experimentally.
With the recent advances in the preparation of nonmaximally entangled 
pure\cite{White:99} and mixed\cite{White:00} polarisation-entangled 
quantum states  we now have a source for which there is a high 
degree of control over the degree of entanglement and purity of the
state. This allows us to consider a wide variety of states and
examine what operations can be performed so as to make the
state more useful in the context of an entanglement resource. 

\begin{center}
\begin{figure}
\epsfig{figure=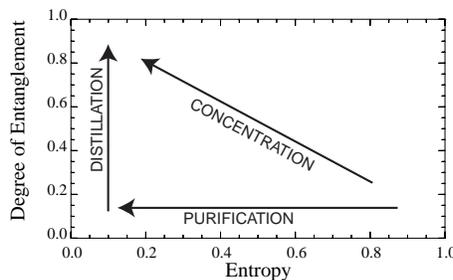,width=60mm}
\caption{\label{fig:ent-pur} A schematic representation of the
entanglement manipulation processes in terms of the degree of
Entanglement and Entropy of a state. We propose this distinction between
distillation, purification and concentration in an attempt to clarify 
terminology in the mixed state domain.}
\end{figure}
\end{center}

For the purposes of describing the possible manipulation of a state 
we will define the following three concepts of distillation, 
purification and concentration (illustrated schematically in 
Figure (\ref{fig:ent-pur})) as follows,
\begin{itemize}
\item \it Distillation\rm: Increasing the entanglement of a state.

\item\it Purification\rm: Increasing the purity of a state 
(decreasing its entropy). This is not
purification with respect to some particular state, for example obtaining a
singlet state from a mixed state.

\item \it Concentration\rm: Increasing both the entanglement and the
           purity of a mixed state.
\end{itemize}
These concepts have been used almost interchangeable in the literature 
but we will follow our primitive definitions to avoid potential confusion.
In this letter it is the concentration of a state that is the main aim 
for the maintenance or recovery of an entanglement resource. 

Let us now specify the measures which we will be using 
to characterise the degree of entanglement and purity of a state.
The entanglement and purity of a state can be determined using distinct
measures. Here we will restrict our attention to $2 \times 2$ systems 
and hence will use analytic expressions for 
The Entanglement of  Formation and Entropy as our respective 
measures. The Entanglement of formation as introduced by Wootters
\cite{Wooters1:98} is found by considering that for a general two
qubit state, $ \rho $, the "spin-flipped state" $\tilde{\rho}$ is given by
\begin{eqnarray}
  \tilde{\rho} = (\sigma _y \otimes \sigma _y) \rho^{*} (\sigma _y \otimes
\sigma _y)
\end{eqnarray}
where $\sigma_y$ is the Pauli operator in the computational basis. We
calculate the square root of the eigenvalues $ \tilde{\lambda}_{i} $ of
$\rho \tilde \rho$, in descending order, to determine the
``Concurrence'',
\begin{eqnarray}
C(\rho) = max\{\tilde{\lambda}_1 - \tilde{\lambda}_2 -\tilde{\lambda}_3
-\tilde{\lambda}_4,0 \} 
\end{eqnarray}
The Entanglement of Formation  (EOF) is then given by
\begin{eqnarray}
 E(C(\rho)) = h\left(\frac{1 + \sqrt{1 - C(\rho)^2}}{2} \right)
\end{eqnarray}
where $ h $ is the binary entropy function
\begin{eqnarray}
h(x) = -x\log(x) - (1 - x)\log(1 - x)
\end{eqnarray}
The entropy of the density matrix $ \rho $ (our purity measure) 
is given by
\begin{eqnarray}
S = -\sum_{i=1}^{4} \lambda_i \log_{4} \lambda_i
\end{eqnarray}
where $ \lambda_i $  are the eigenvalues of $ \rho $.

\begin{center}
\begin{figure}
\epsfig{figure=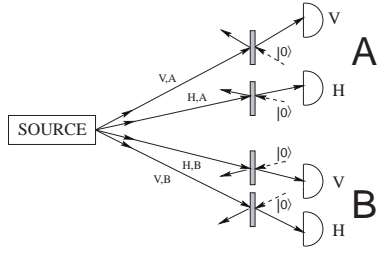,width=50mm}\\
\caption{\label{fig:exp2} The schematic model of the BS concentration
protocol. The source produces an initial state that can be controlled
in terms of the entanglement and purity and is thus able to provide a
range of initial states for manipulation. This state is then spatially
separated with respect to its polarisation modes and then incident on 
separate beam splitters, BSs, with a different variable reflectivity. 
By \it tuning \rm the variable BS it is possible to obtain the desired output
state with the corresponding 
coincidence detections for A and B.}
\end{figure}
\end{center}

We will now describe our entanglement manipulation protocol and emphasise its
simplicity. The experimental arrangement for our 
protocol is  described by the schematic in figure(\ref{fig:exp2}). 
The aim of our protocol is to manipulate mixed states and
enhance their degree of entanglement. Let us consider an initial 
state composed of two subsystems, A and B, 
each represented by a general $2 \times 2$
matrix. We will describe the joint state of the system, $AB$, 
in the polarisation basis, $\{|VV\rangle,|VH\rangle, |HV\rangle, 
|HH\rangle\}$, as
\begin{eqnarray}
\hat{\rho}_{ABin} &=& \left( \begin{array}{cccc}
		\rho_{11} & \rho_{12} & \rho_{13} & \rho_{14} \\
	        \rho_{12}^* & \rho_{22} & \rho_{23} & \rho_{24} \\
		\rho_{13}^* & \rho_{23}^* & \rho_{33} & \rho_{34} \\
	        \rho_{14}^* & \rho_{24}^* & \rho_{34}^* & \rho_{44} \\ \end{array} \right)
\end{eqnarray}
with the $\hat{\rho}_{ij}$ satisfying the requirements for a legitimate
density matrix. From our source (see figure(\ref{fig:exp2})) 
we have four polarisation modes (two for A and two for 
B). These polarisation modes are spatially separated and 
input onto beam splitters (BS), with independent and variable 
transmission coefficients. The second input port of each of these 
beam splitters are assumed to be vacuums. With perfectly efficient 
photodetectors it would be possible to monitor the second output mode 
of each of these beamsplitter and use the results to conditionally 
select the concentrated state we wish to produce. We know that if the detection 
of a photon is made in any of the second output ports then the preparation 
process is considered to have failed. Non-detection (with perfectly 
efficient detectors) at all the second output ports is required to 
prepare our state and here is the problem with current single photon detection 
efficiencies. Photon detectors have a finite efficiency and it 
possible that a photon present at these second output ports will not 
be detected. Hence we will not get the conditioned state we desire.
Instead we will examine the transmitted modes of the beamsplitter 
and consider the situations where joint 
coincidences are registered at the photodetectors 
of the two subsystems A and B, or Alice and Bob if you prefer. While 
this is a post selective process it has the advantage that poor 
detection efficiency only decreases the coincidence count rate. As we 
discard any information present at the second output of the 
beamsplitters, the protocol we describe is not unitary. 

If we consider that having each mode incident on a BS has the
effect of expanding the Hilbert space of the system, then in the
expanded Hilbert space we can manipulate the state and then project 
it back onto the polarisation coincidence basis. The BSs transform each mode in 
the following way
\begin{eqnarray}
 |V,H \rangle |0 \rangle \rightarrow \eta_{v,h}|V,H \rangle |0 \rangle +
  \sqrt{(1-\eta_{v,h}^2)}|0\rangle |1\rangle
\end{eqnarray}
and hence we obtain an output density matrix for this reduced system of the form
\begin{eqnarray}
\hat{\rho}_{ABout} &=&{\cal N} \left( \begin{array}{cccc}
\rho_{11}\eta_{va}^2\eta_{vb}^2&\rho_{12}\eta_{va}^2\eta_{vb}\eta_{hb}& \rho_{13}\eta_{va}\eta_{ha}\eta_{vb}^2 & \rho_{14}\eta \\
\rho_{12}^*\eta_{va}^2\eta_{vb}\eta_{hb}&\rho_{22}\eta_{va}^2\eta_{hb}^2&\rho_{23}\eta&\rho_{24}\eta_{va}\eta_{ha}\eta_{hb}^2 \\
\rho_{13}^*\eta_{va}\eta_{ha}&\rho_{23}^*\eta&\rho_{33}\eta_{ha}^2\eta_{vb}^2&\rho_{34}\eta_{ha}^2\eta_{vb}\eta_{hb} \\
\rho_{14}^*\eta&\rho_{24}^*\eta_{va}\eta_{ha}\eta_{hb}^2&\rho_{34}^*\eta_{ha}^2\eta_{vb}\eta_{hb} &\rho_{44}\eta_{ha}^2\eta_{hb}^2 \\
		\end{array} 
\right)
\end{eqnarray}
where $\eta = \eta_{va}\eta_{ha}\eta_{vb}\eta_{hb}$ and
$\eta_{v,h|a,b} $ are the vertical and horizontal polarisation 
transmission coefficients for subsystems A and B. The normalisation is given by
\begin{eqnarray}
{\cal N} = [\rho_{11}\eta_{va}^2\eta_{vb}^2 + \rho_{22}\eta_{va}^2\eta_{hb}^2 + \rho_{33}\eta_{ha}^2\eta_{vb}^2 + \rho_{44}\eta_{ha}^2\eta_{hb}^2]^{-1}
\end{eqnarray}
and the probability of obtaining the desired output state is determined from
the trace of the unnormalized BS-transformed density matrix, ${\cal
N}^{-1}$, and thus is dependent on the transmission coefficients. This is the probability of obtaining the output state
once the BS parameters have been determined.

This scheme is more easily understood by considering the behaviour
of pure states under the protocol.  As such we now illustrate the 
distillation process with a specific example. We will examine a non-maximally
entangled pure state and show how to recover a maximally entangled
state via our protocol. Consider an initial state produced by our source of the form
\begin{eqnarray}\label{eq:nmeps1}
|\varphi_{in} \rangle_{ab} = {\cal N}_{1}[\epsilon_1|VV\rangle_{ab} + \epsilon_2e^{i\phi}|HH\rangle_{ab}]
\end{eqnarray}
or alternatively
\begin{eqnarray}\label{eq:nmeps2}
|\varphi_{in} \rangle_{ab} = {\cal N}_{1}[\epsilon_1|VH\rangle_{ab} + \epsilon_2e^{i\phi}|HV\rangle_{ab}]
\end{eqnarray}
where
\begin{eqnarray}
{\cal N}_{1}^2=[|\epsilon_1|^2 + |\epsilon_2|^2]^{-1}
\end{eqnarray}
Assuming the polarisation modes are all spatially separated we input
them onto separate BSs (see figure (\ref{fig:exp2})). We can choose
to manipulate the BSs at A and B independently to find the optimal
output for a given state. For convenience we consider a state of the form of 
(\ref{eq:nmeps1}) which allows us to simplify the analysis. With
this in mind we can set $\eta_{va} = \eta_{vb} = \eta_v$ and
$\eta_{ha} = \eta_{hb} = \eta_b$.

The state of our system after the  BSs (assuming vacuum
inputs to the second BS ports) is
\begin{eqnarray}
|\varphi_{total} \rangle_{AB} &=&{\cal N}_{1} \left[
\epsilon_1\eta_v^2|VV\rangle_{AB}|00\rangle +
\epsilon_2e^{i\phi}\eta_h^2|HH\rangle_{AB} |00\rangle \right. \nonumber \\
&\;&\;\;\;+\epsilon_1\eta_v \sqrt{(1-\eta_{v}^2)}\left\{
|V0\rangle_{AB}|01\rangle+
|0V\rangle_{AB}|10\rangle \right\} \nonumber \\
&\;&\;\;\;+\epsilon_2e^{i\phi}\eta_h \sqrt{(1-\eta_{h}^2)}\left\{
|H0\rangle_{AB}|01\rangle+
|0H\rangle_{AB}|10\rangle  \right\} \nonumber \\
&\;&\;\;\; + \epsilon_1 \left(1-\eta_v^2\right)|00\rangle_{AB}|11\rangle
\nonumber \\
&\;&\;\;\;\left. + \epsilon_2e^{i\phi}\left(1-\eta_h^2\right)|00\rangle_{AB}
|11\rangle\right]
\end{eqnarray}
The outcomes we are interested in are in the joint coincidence basis of A,B and
hence the vacuum state components are removed from consideration
leaving an effective output state of the form 
\begin{eqnarray}
|\varphi_{out} \rangle_{AB} = {\cal
N}_{2}[\epsilon_1\eta_v^2|VV\rangle_{AB} +
\epsilon_2e^{i\phi}\eta_h^2|HH\rangle_{AB}]
\end{eqnarray}
where the normalisation in this coincidence basis is
\begin{eqnarray}
{\cal N}_{2}^2 = [|\epsilon_1|^2\eta_v^4 + |\epsilon_2|^2\eta_h^4]^{-1}
\end{eqnarray}
For maximal entanglement we have the following simple relationship
\begin{eqnarray}
|\epsilon_1| \eta_v^2 = |\epsilon_2| \eta_h^2
\end{eqnarray}
We observe that the entanglement of the output state is dependent
on the transmission coefficients of the BSs. Further, this protocol
can always take a non-maximally entangled state and obtain a pure 
maximally entangled one. This protocol can also incorporate a phase
adjuster at either A or B to tune any relative phase difference for
the state. If we had considered states of the form of
(\ref{eq:nmeps2}) then we would need to consider the tuning
parameters independently such that the requirement for a pure
maximally entangled state is then
\begin{eqnarray}
|\epsilon_1| \eta_{va}\eta_{hb} = |\epsilon_2| \eta_{vb}\eta_{ha}
\end{eqnarray}
This is where the protocol differs from the Procrustean method of
Bennett \it et.al \rm \cite{Bennett3:96}. We have introduced individual
depolarising channels, thus obtaining more degrees of freedom, and
so allowing the protocol to be extended to mixed states. It is
important to mention again that with perfect single photon detection
it is possible to monitor the discarded ports for each of the modes,
thus preparing the desired state by conditioned measurements.

Let us now turn our attention to the concentration of mixed
states. As an extension to the distillation process we take the 
density matrix $\hat{\rho}_{ABin}$ to be a mixture of the density
matrices of two of the Bell-type states, (\ref{eq:nmeps1}) and 
(\ref{eq:nmeps2}), one of which, say (\ref{eq:nmeps1}), is maximally 
entangled, $\epsilon_1 = \epsilon_2 = 1$. The mixing can be controlled
by the parameter $\gamma$, that is,
\begin{eqnarray}\label{eq:rhoineg}
\hat{\rho}_{ABin} &=& \gamma {\cal N}_{1}^{2} \left( \begin{array}{cccc}
		|\epsilon_1|^2 & 0 & 0 & \epsilon_1^{*}\epsilon_2 \\
	        0 & 0 & 0 & 0 \\
		0 & 0 & 0 & 0 \\
	 \epsilon_1\epsilon_2^{*} & 0 & 0 & |\epsilon_2|^2 \\ \end{array} \right)
 +  \frac{1-\gamma}{2}\left( \begin{array}{cccc}
		0 & 0 & 0 & 0 \\
	        0 & 1 & 1 & 0 \\
		0 & 1 & 1 & 0 \\
	 	0 & 0 & 0 & 0 \\ \end{array} \right)
\end{eqnarray}
This state is one of many of the range of mixed states that can be
concentrated and has been chosen to easily show the protocols
extension from pure to mixed states, from distillation to concentration.
Using the BS protocol illustrated in figure(\ref{fig:exp2}) the
output state for (\ref{eq:rhoineg}) in the coincidence basis, $AB$, can
be represented as
\begin{eqnarray}
\hat{\rho}_{ABout}&=&{\cal N}_{3}^{2} \left( \begin{array}{cccc}
\gamma|\epsilon_1|^2\eta_{va}^2\eta_{vb}^2&0&0&\gamma \epsilon_1\epsilon_2^{*}\eta  \\
 0 &\Gamma \eta_{va}^2\eta_{hb}^2 &\Gamma \eta & 0 \\
0 & \Gamma \eta  & \Gamma \eta_{ha}^2\eta_{vb}^2 & 0 \\
\gamma \epsilon_1^{*}\epsilon_2\eta & 0 & 0 &\gamma |\epsilon_2|^2\eta_{ha}^2\eta_{hb}^2 \\ \end{array} \right)
\end{eqnarray}
with $\Gamma = \frac{(1-\gamma)}{2}$
and the normalisation ${\cal N}_{3}$  given by
\begin{eqnarray}
{\cal N}_{3}^{2} =
[\gamma(|\epsilon_1|^2\eta_{va}^2\eta_{vb}^2+|\epsilon_2^2|\eta_{ha}^2\eta_{hb}^2)+\Gamma(\eta_{va}^2\eta_{h_{b}}^2 + \eta_{ha}^2\eta_{vb}^2)]^{-1}
\end{eqnarray}

\begin{center}
\begin{figure}
\epsfig{figure=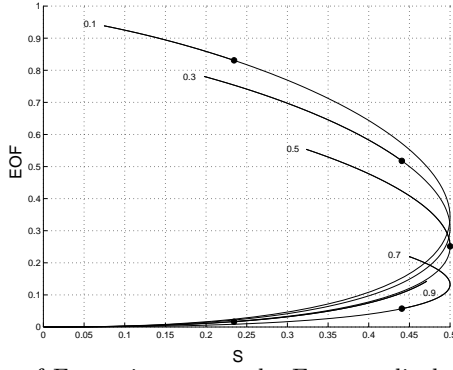,width=60mm}
\caption{\label{fig:conc}  The plot of the Entanglement of Formation 
versus the Entropy displays the concentration characteristics of the 
idealised schematic of figure(\ref{fig:ent-pur}). We consider states of the 
form of (\ref{eq:rhoineg})
with $(\epsilon_1,\epsilon_2) = (1,0.1)$, the values for $\gamma$ are
labeled at the peak of each curve. The curves illustrate  the entire
range of values a state can attain in the (S,EOF) plane  as the BS
parameters, $\eta_{va} = \eta_{vb}=\eta_v$, are varied,
$0<\eta_{v}<1$. The initial states are marked as filled black
circles. If we consider the state $\gamma = 0.1$, initially (S,EOF) = 
(0.23,0.84), then as we decrease $\eta_v$ from an initial value of 1 we 
approach the maximum concentration value (S,EOF) = (0.075,0.94) at
$\eta_v= 0.32$. If we continue to decrease $\eta_v$ we then follow the
curve back down through our initial point on the plane and from then on the
state deteriorates from its original value. We also note that the
other curves have similar concentration characteristics for $\gamma
\le 0.5$. $\gamma = 0.5$ corresponds to the case where the two
Bell-type states are evenly mixed.}
\end{figure}
\end{center}

In figure (\ref{fig:conc}) we display the effect of our protocol for 
a range of $\gamma$ values with $\epsilon_1=1$ and $\epsilon_2=0.1$ 
(the $\gamma$  values are labeled at the peak of each curve). The initial points for the fixed  $\gamma$, $\epsilon_1$ and $\epsilon_2$ 
are displayed as solid dots. These curves represent the behaviour
of the Entropy and EOF of the states as the BSs are tuned to optimise
both. We see how this class of state can be improved is dependent on the
amount mixing. The behaviour of the state is similarly dependent
on the degree of entanglement in the pure state components of the
mixed state of (\ref{eq:rhoineg}), variations in $\epsilon_{1,2}$,
though this is not explicitly shown here.

The curves in figure (\ref{fig:conc}) represent the range of (S,EOF)
values for the output states from our protocol. We take the
specific case of $\gamma = 0.1$ and observe the variation of (S,EOF) as we tune
$\eta_{va} = \eta_{vb} = \eta_v  $. From the initial state marked with a
black circle at (S,EOF) = (0.23,0.84) with $\eta_v=1$ we then adjust
the BSs, moving up the curve, to a state with (S,EOF) =
(0.075,0.94) for $\eta_v= 0.32$. This constitutes a turning point on the plane and if we
continue decreasing $\eta_v$ we follow the curve back to our initial point in the plane after which
the entanglement-entropy properties of the state deteriorate from the
original values.

What does the state look like? We observe that with 
($\epsilon_1,\epsilon_2,\gamma) = (1.00,0.10,0.30)$ and allowing all
the light through the horizontal BS (an optimal setting provided 
$|\epsilon_1| > |\epsilon_2|$ to maximise the output), and tuning the 
vertical beam splitters transmission to $\eta_{v} = 0.32$ we can take 
an initial state
\begin{eqnarray}\label{eqn:rhoinsp}
\hat{\rho}_{ABin} = \left( \begin{array}{cccc}
		0.297 & 0 & 0 & 0.030 \\
	        0 & 0.350 & 0.350 & 0 \\
		0 & 0.350 & 0.350 & 0 \\
	        0.030 & 0 & 0 & 0.003 \\ \end{array} \right)
\end{eqnarray}
to an output state
\begin{eqnarray}
\hat{\rho}_{ABout} = \left( \begin{array}{cccc}
		     	 0.039 & 0 & 0 & 0.039 \\
	       		 0 & 0.461 & 0.461 & 0 \\
		     	 0 & 0.461 & 0.461 & 0 \\
	       		 0.039 & 0 & 0 & 0.039 \\ \end{array} \right)
\end{eqnarray}
This output state has an increase in the Entanglement of Formation
from EOF = 0.52 to EOF = 0.78, while the entropy of the system has
decreased from S = 0.30 to S = 0.20, this result is achieved with a 
finite probability P = 7.6\%. 

There exists a critical point with respect to
concentration at $\gamma = 0.5$ which corresponds to the case
where the two pure states of (\ref{eq:rhoineg}) are evenly mixed. For
those states with the mixing parameter $\gamma \le 0.5$ concentration
is possible whilst for those states above this value the entanglement
can be increased but this is at the cost of purity. All of these
states can be concentrated if we choose to tune another BS, thus
highlighting the need for all four BSs. Similarly if we considered a
mixture of the pure states of (\ref{eq:nmeps1}) and (\ref{eq:nmeps2}), 
where both had $\epsilon_{1,2} \ne 1$, then we find that concentration
is still achievable. 

Now let us consider the incoherent sum of a pure state and a mixed
state and take as an example of this the Werner state, a mixture of
the identity and some fraction of a pure state. If the pure state
fraction of the Werner state is a non-maximally entangled pure state,
then it is possible to increase the entanglement of the state. However 
this entanglement increase comes at the cost of purity and is bound by 
the amount of entanglement that would be inherent in a Werner state using a
maximally entangled pure state. 

\label{CONCLUSION}

In conclusion, we have proposed an entanglement concentration protocol that is
experimentally realisable and can produce a finite concentration of Bell pairs
from some initially mixed states. The key point here is that whilst this
is achievable we are more interested in the entanglement properties
then the final form of the state. Indeed  with such a simple protocol the range of possible
tests with respect to quantum information and entanglement are quite
diverse, and whilst this protocol does require some knowledge of the
state in determining the tuning parameters and is a non-unitary
operation, we believe it should provide a most useful tool in the
exploration of mixed state entanglement.\\

 The authors would like to thank A.G. White and P.G. Kwiat for useful 
discussions with respect to the practicality of the experimental
implementation of this scheme. WJM would like to acknowledge the support of the Australian
Research Council.


\end{document}